\documentclass[times,authoryear]{elsarticle}

\usepackage[utf8]{inputenc}
\usepackage{graphicx}
\usepackage{url}
\graphicspath{ {images/} }
\usepackage{caption}
\usepackage{amsmath}
\usepackage{verbatim}

\usepackage[authoryear]{natbib}

\usepackage{listings}
\usepackage{subfig}
\usepackage{fancyvrb}

\usepackage{lineno}

\def\tsc#1{\csdef{#1}{\textsc{\lowercase{#1}}\xspace}}
\tsc{WGM}
\tsc{QE}
\tsc{EP}
\tsc{PMS}
\tsc{BEC}
\tsc{DE}

\usepackage{hyperref}
\hypersetup{
    colorlinks=true,
    linkcolor=blue,
    filecolor=magenta,      
    urlcolor=cyan,
    pdftitle={Overleaf Example},
    pdfpagemode=FullScreen,
    }

\usepackage{arydshln}

\begin{document}

\title{Analyze Mass Spectrometry data with Artificial Intelligence to assist the understanding of past habitability of Mars and provide insights for future missions}

\date{September 2023}

\author[1]{Ioannis Nasios}

\ead{ioannis.nasios@nodalpoint.com}

\affiliation[1]{organization={Nodalpoint Systems},
    addressline={Mitropoleos 43}, 
    city={Athens},
    postcode={105 56}, 
    country={Greece}}

\begin{frontmatter}

\begin{abstract}

This paper presents an application of artificial intelligence on mass spectrometry data for detecting habitability potential of ancient Mars. Although data was collected for planet Mars the same approach can be replicated for any terrestrial object of our solar system. Furthermore, proposed methodology can be adapted to any domain that uses mass spectrometry. This research is focused in data analysis of two mass spectrometry techniques, evolved gas analysis (EGA-MS) and gas chromatography (GC-MS), which are used to identify specific chemical compounds in geological material samples. The study demonstrates the applicability of EGA-MS and GC-MS data to extra-terrestrial material analysis. Most important features of proposed methodology includes square root transformation of mass spectrometry values, conversion of raw data to 2D sprectrograms and utilization of specific machine learning models and techniques to avoid overfitting on relative small datasets. Both EGA-MS and GC-MS datasets come from NASA and two machine learning competitions that the author participated and exploited. Complete running code for the GC-MS dataset/competition is available at GitHub.\footnote{\url{https://github.com/IoannisNasios/MarsSpectrometry2_GasChromatography}} Raw training mass spectrometry data include [0,1] labels of specific chemical compounds, selected to provide valuable insights and contribute to our understanding of the potential past habitability of Mars.

\end{abstract}

\begin{keyword}
Mars; Mars, surface; Terrestrial planets; Spectroscopy;  
\end{keyword}

\end{frontmatter}

\section{Introduction}
\label{sec:introduction}
Machine learning is used in more and more applications and domains. Starting from a labeled training dataset, machine learning models (algorithms) can be trained, can learn the dataset. When model training is over it's hyperparameters are adjusted. Trained models can then be stored and loaded for inference on any data alike. Machine learning models can also be used for learning mass spectrometry data of material samples. Detection of presence or absence of specific chemical compounds in geological samples can be utilized in many different applications. Compounds that need to be detected are driven by the issue that has to be addressed every time. Presented methodologies here, although are tuned for specific datasets, can also be used for other mass spectrometry datasets.

Raw data for this research was provided by NASA. Whether selected chemical compounds could all be produced by abiotic processes or provide evidence of life is not a concern of this manuscript. Present methodology could be adapted to any other dataset. Including new compound(s) or excluding existing compound(s) that could better explain habitability conditions or distinguish abiotic processes from life, would improve the study for past habitability of Mars.

AI/ML significance for space data treatment is highly important. This is clearly imprinted in the Artificial Intelligence for the advancement of Lunar and Planetary Science and Exploration of the Astrobiology Decadal Survey 2023-2032 whitepaper, \citep{varatharajan2021artificial} and therein references. As study concluded, AI-driven tools and methodologies have the potential to support the lunar and planetary science community, effectively exploit the currently available datasets, and help prepare for the next decade of science and exploration.  In addition, \citep{slingerland2022adapting} provided a set of best practices to assist the study of an AI-based autonomy that maximizes trust and lowers the barriers to mission adoption.

Mass spectrometry instruments are crucial tools of expeditions seeking potential signs of habitability or even for detecting indications of life on celestial objects. For this, NASA conducted two machine learning competitions to not only aid researchers in hastening their analysis of data, but also demonstrate the feasibility of applying data science and machine learning approaches to intricate mass spectrometry data in forthcoming missions. The goal was to detect the presence of certain families of chemical compounds in geological material samples using evolved gas analysis (EGA-MS) or gas chromatography (GC-MS) mass spectrometry data, collected for Mars exploration missions. These families are of rocks, minerals, and ionic compounds relevant to understanding Mars past habitability potentials. These research competitions-datasets aim as stated, to \emph{"guide science operations, reduce reliance on ground-in-the-loop analysis, and prioritize transmission over long distances"}. With extremely limited communication between Mars and Earth in both time availability and bandwidth, advanced autonomy must be incorporated into future rovers, \citep{ono2022machine}. 
To a future mission on a terrestrial body a rover could be programmed to use these analysis to take further actions such as resampling a spot without waiting to receive an explicit command. This would be of more value for more distant objects due to the larger communication time delay.

Many missions have been sent on Mars and more are planned for the near future. Among these missions there is the Curiosity rover which is equipped with mass spectrometers for geological sample analysis. The Curiosity rover which has been on Mars since August 5 2012, has proved to be highly productive. Onboard Curiosity there is the Sample Analysis at Mars instrument \href{https://mars.nasa.gov/msl/spacecraft/instruments/sam/}{(SAM)} 
which analyzes geological samples using mass spectrometry methods. SAM's instruments have produced extensive scientific research such as the evolved gases analysis for the understanding of the oxidation state of sulfur in the samples from the clay-bearing Glen Torridon region, \citep{wong2022oxidized} and the identification of organic molecules by gas chromatography, \citep{millan2016situ}. Curiosity is still active and it is unknown exactly when it will be disabled. Although presented methodology and data are for SAM's instruments, it is of more importance that could provide valuable insights for future missions containing instruments alike.

Several Mars missions have utilized various spectrometry instruments to study the Martian environment and assess its habitability. Many researchers have analyzed and experimented with these data. Spectrometry types used for Mars include the SuperCam infrared spectrometer on the Perseverance rover which is designed to analyze the mineralogy and organic compounds in rock and soil samples \citep{fouchet2022supercam}, the SPICAM UV spectrometer on the Mars Express orbiter which focuses on studying the Martian atmosphere \citep{bertaux2006spicam} and other spectroscopy techniques such as Raman spectroscopy and laser-induced breakdown spectroscopy  \citep{clegg2014planetary}. 

Evolved gas analysis mass spectrometry is a powerful technique used for the characterization of various types of samples, including extra-terrestrial materials, soils, minerals and organic compounds. This method involves the controlled heating (temperatures are recorded) of a sample in a gas flow and the analysis of the evolved gases using mass spectrometry to identify and quantify the chemical species released. Recent studies have demonstrated the applicability of EGA-MS to a wide range of fields, from in situ resource utilization water extraction studies on Mars simulant soils \citep{clark2020jsc}, to the speciation of minor minerals in clays \citep{zumaquero2020application} and the evaluation of synergistic effects and kinetic data in co-pyrolysis of wood and plastic \citep{nardella2021co}. Also, \cite{verchovsky2020quantitative} proposed a quantitative approach for the analysis of extra-terrestrial samples, highlighting the importance of a thorough understanding of the gas release mechanisms and their dependence on the sample characteristics. Here, state-of-the-art machine learning models and techniques are used for the classification of various types of Martian analogue samples. This could also be used to gain insights into the potential use of mass spectrometry for the exploration of other planets and moons in the solar system, and for the analysis of samples returned from space missions.

Gas chromatography mass spectrometry is also a powerful technique widely used in various fields, including food science, chemistry, environmental analysis and pharmaceuticals. The combination of GC-MS and machine learning provides fast identification of complex mixtures. \cite{chou2021planetary} examined several possible approaches to agnostic life detection using mass spectrometry. The application of GC-MS includes the classification of food products. \cite{tan2018determining} developed an artificial neural network (ANN)-based electronic nose system coupled with GC-MS to determine the degree of roasting in cocoa beans, while \cite{pastor2022classification} utilized GC-MS liposoluble fingerprints and automated machine learning to classify cereal flour samples. \cite{aghili2022detection} also used GC-MS to detect fraudulent activity in sesame oil by combining artificial intelligence with chemometric methods and chemical compound characterization. This demonstrates the versatility of GC-MS in providing precise, sensitive and accurate analysis in various fields, which combined with artificial intelligence can make an essential analytical tool for researchers and industries alike.

\section{Material and methods}
\label{sec:Material and methods}
This research comes after author's participation in both NASA's machine learning competitions, \href{https://www.drivendata.org/competitions/93/nasa-mars-spectrometry/}{"Mars Spectrometry: Detect Evidence for Past Habitability"} (7th place out of 713 participants, Feb-April 2022)  and \href{https://www.drivendata.org/competitions/97/nasa-mars-gcms/}{"Mars Spectrometry 2: Gas Chromatography"} (3rd prize winner, Sep-Oct 2022), on drivendata platform, as a validation of machine learning capabilities on mass spectrometry data for compounds classification. Experience gained at the first competition applied on the second one successfully. Working with both types of data, share similarities and can give a better and wider aspect of understanding on how mass spectrometry data and artificial intelligence models can be used effectively.

\subsection{Prepare datasets for models}
\label{sec:Prepare datasets}
Raw data comes as one csv file per sample. For the EGA-MS data, below we can see the beginning of a sample file. For GC-MS, there is no temperature column and instead of abundance there is an equivalent intensity column. The main difference between the two datasets is the temperature dimension which doesn't exist for the GC-MS data. As stated for the GC-MS \emph{"Time can be used as a proxy for temperature, but the temperature ramp is not exactly known nor the same across samples. It is always the case that observations at later times are of compounds that were released at higher temperatures. In most samples, it is expected that the temperature remains constant for the first 0-5 minutes, and then increases at an approximate rate of 5-10 degrees per minute until it reaches about 300 degrees. However, because the time before the ramp begins and the rate at which the temperature increases changes across samples, the same time will represent different temperatures across samples"}. Temperature recordings in EGA-MS can provide a more detailed knowledge and understanding of when and how much gas is released from the sample.

\begin{Verbatim}[fontsize=\footnotesize, frame=lines, label=Raw data file sample (EGA-MS),commandchars=\\\{\}]

 	 time        temp 	  \emph{m/z} 	abundance
0 	0.0 	32.226 	0.0 	7.020079e-10
1 	0.0 	32.226 	1.0 	1.045714e-09
2 	0.0 	32.226 	2.0 	3.802511e-10
3 	0.0 	32.226 	3.0 	5.783763e-10
4 	0.0 	32.226 	4.0 	7.238828e-08
...       ... 	... 	   ... 	...

\end{Verbatim}

\subsubsection{Dataset Labels}
\label{sec:Dataset target}
For EGA-MS data there are 10 classes while for the GC-MS  there are 9 classes, each indicating presence of material belonging to the respective rock, mineral, or organic compound families, in the sample. As these are multilabel tasks, there can be more than one classes present in one sample or there can be samples without any class at all. A detailed account for both datasets can be seen in \autoref{tab:data_classes}.

\begin{table}[ht]
\caption{Classes and number of samples for each dataset}\label{tab:data_classes}
\centering
\begin{tabular} {l l  l  l l}
\hline 
 & \textbf{EGA-MS} & & \textbf{GC-MS} & \\
\hline 
1 & Basalt          & 122 &  Alcohol  & 34 \\
2 & Carbonate       & 133 &  Aromatic  & 103\\
3 & Chloride        & 132 &  Carboxylic\_Acid & 165\\
4 & Iron Oxide      & 232 &  Chlorine\_bearing\_compound & 26 \\
5 & Oxalate         & 43  &  Hydrocarbon & 307 \\
6 & Oxychlorine     & 240 &  Mineral  & 180 \\
7 & Phyllosilicate  & 343 &  Nitrogen\_bearing\_compound & 139 \\
8 & Silicate        & 140 &  Other\_Oxygen\_bearing\_comp. & 26 \\
9 & Sulfate         & 211 &  Sulfur\_bearing\_compound  & 36  \\
10 & Sulfide         & 496 & &  \\ 
\hline 
Sum             &  & 1645 &  & 1016 \\
Num. Samples     &  & 1059 &   & 1121 \\
with NO class   &   & 139 & & 450 \\
with \textgreater1 classes  &   & 340 &  & 178 \\
\hline 

\end{tabular}
\end{table}

\subsubsection{EGA-MS}
\label{sec:EGAMS_dataset}
\begin{figure}[h]
\centering
\begin{tabular}{@{}c@{}}
   \includegraphics[width=0.99\linewidth,trim=0 0 0 45, clip]{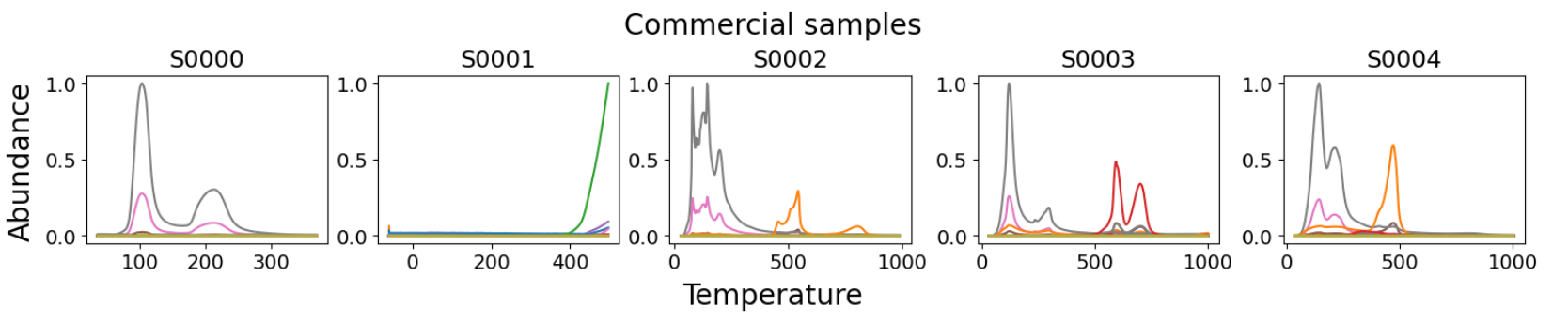}\\[\abovecaptionskip]
   \small (a) 
\end{tabular}   
\begin{tabular}{@{}c@{}}
   \includegraphics[width=0.99\linewidth,trim=0 0 0 45, clip]{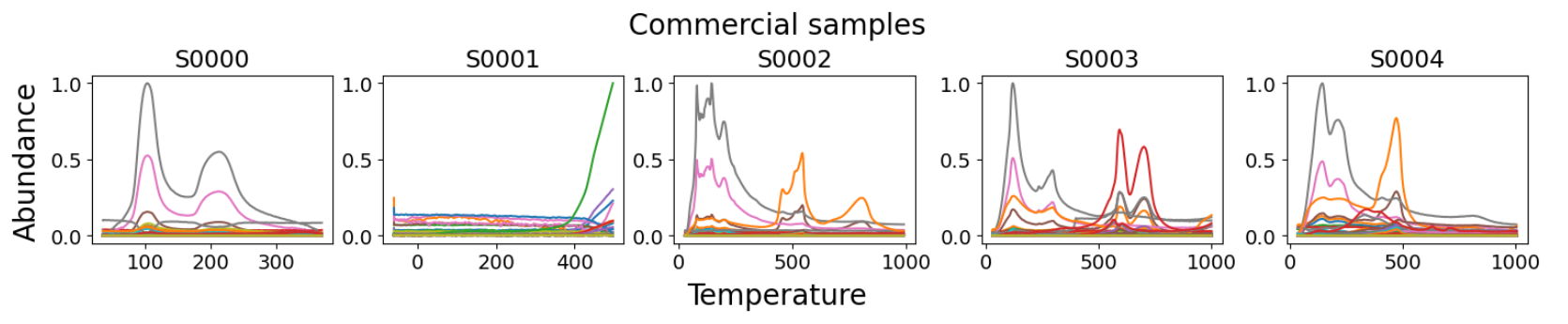}\\[\abovecaptionskip]
   \small (b) 
\end{tabular}      
\caption[ion plots]{Every plot have 100 ion - lines. Temperature on horizontal axis and ions abundance on vertical. (a) Commercial samples. (b) Same Commercial samples with root transformation abundance.}
\label{fig:ega_root_noroot}%
\end{figure}

\begin{figure}[h]
    \centering
    \includegraphics[width=0.99\textwidth,trim=0 0 0 40, clip]{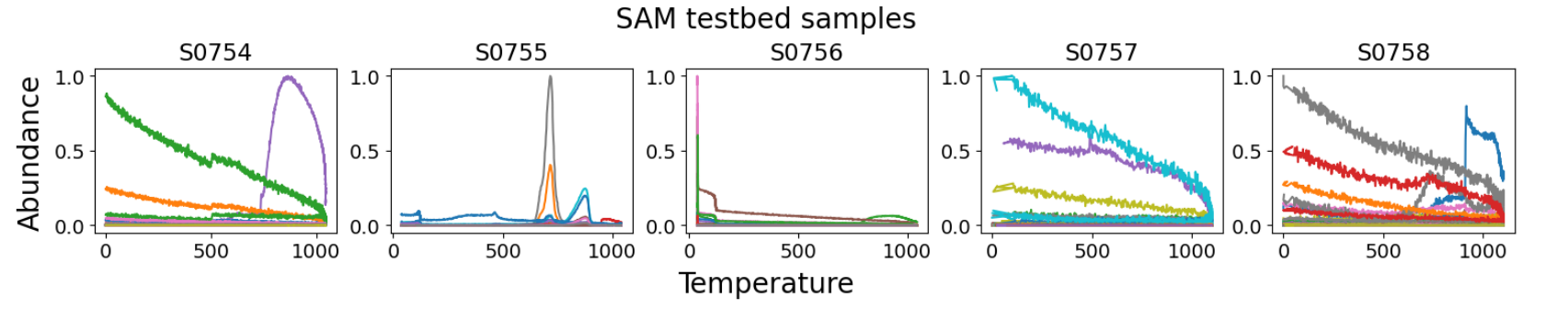}
    \caption{SAM testbed samples}
    \label{fig:index_sam}
\end{figure}

Training data consists of 1047 samples from commercial instruments and 12 samples from SAM testbed replica instrument. Building a modelling pipeline that performs well also on SAM testbed samples added extra difficulty and complexity. As can be observed in plots of \autoref{fig:ega_root_noroot}(a) and \autoref{fig:index_sam}, and a more thorough data exploration, our two sample types are quite different. There seems to be more noise and higher temperatures on the SAM samples. Test data includes 64 SAM samples and 446 commercial, a higher ratio of SAM samples over commercial compared to the training data. Finally, there are also some supplemental data which includes 41 SAM and 220 commercial samples (subset with "he" carrier gas and NO different pressure) but these data contain no labels and are only used in an unsupervised way (pseudo-labeling).

The data has been collected from multiple labs from NASA's Goddard Space Flight Center and Johnson Space Center. Two kinds of instruments were used to conduct the measurements:

\begin{enumerate}
\item Commercial instruments — the data comes from commercially manufactured instruments that have been configured as SAM analogues at the Goddard and Johnson labs. Data collected as whole number of \emph{m/z} values ranging from 0.0 to 100.0.
\item SAM testbed — the data comes from the SAM testbed at Goddard, a replica of the SAM instrument suite on Curiosity. Data collected for \emph{m/z} values up to 534.0 or 537.0 and sometimes included fractional values.
\end{enumerate}

Preparing the data for the models, the following steps are taken:

\begin{enumerate}
\item 
100 first \emph{m/z} (mass-to-charge ratio) values used for a uniform dataset driven by commercial instruments limited extent. 
\item Fractional \emph{m/z} values are removed when present.
\item \emph{m/z 4}, which corresponds to helium carrier gas is also dropped. 
\item For every ion in sample the minimum abundance value is subtracted as this represents the background presence of the ion. This can happen for various reasons, such as contamination from the atmosphere. The new minimum for every ion in sample is zero. (step suggested by competition hosts, no experiment was conducted otherwise).
\item Square root transformation is applied. Experiments conducted such as not using any transformation or using log transformation instead gave worse results.
\item Abundances are normalized from 0 to 1 within a single sample. Normalization almost every time and here to, is an important element of neural networks performance. With this scaling, the relative abundance between samples is lost but what is more important is that the relative abundance between ions of the same sample is preserved.
\end{enumerate}

For primary dataset creation, overall temperature range into bins (of X degrees) and maximum relative abundance calculation within that temperature bin for each \emph{m/z} value are obtained. With 4 different bin temperature widths [50, 100, 200, 400], 4 primary datasets are created. Within each bin the maximum abundance value is acquired. A secondary dataset is also created containing 7 features other than abundance: maximum time, standard deviation of time, maximum temperature, median temperature, standard deviation of temperature and number of records in file.

\subsubsection{GC-MS}
\label{sec:GCMS_dataset}
\begin{figure}[h]
    \centering
    \includegraphics[width=0.9\textwidth,trim=0 0 0 30, clip]{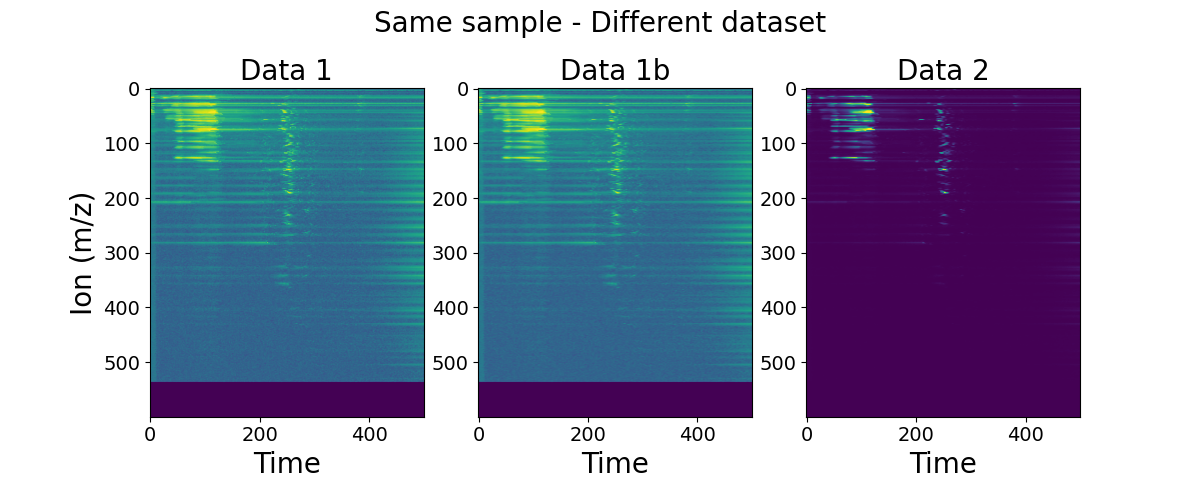}
    \caption{Spectrograms of the same sample for all 3 different initial GC-MS datasets.}
    \label{fig:SameSample_DifferentDataset}
\end{figure}

This dataset was collected only by commercially manufactured instruments that have been configured as SAM analogues at Goddard. For training there are 1121 samples and for test 463. From raw csv files, initially 3 sets of train and test datasets are created. The shape of these datasets is [number of samples, 600 \emph{m/z}, 500 timesteps]. Each one of these samples in created dataset can be viewed as an image, as a spectrogram. For creating these datasets a set of processing steps is required. First, all float raw \emph{m/z} values are rounded to integers. As number of different \emph{m/z} per sample vary, all intensity gaps are zero filled (zero padded at highest or sometimes lowest \emph{m/z} values, visualized as a bottom dark stripe at \autoref{fig:SameSample_DifferentDataset}). As number of timesteps per sample vary, if greater than 500 then use the max in between value else repeat following or previous value. 

\begin{figure}[h]
    \centering
    \includegraphics[width=0.95\textwidth]{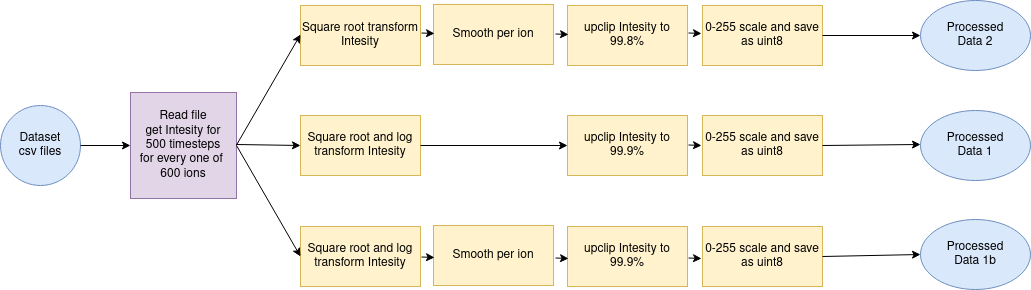}
    \caption{From raw files to initial datasets for GC-MS data}
    \label{fig:create_datasets}
\end{figure}

Differences between all three initial datasets are depicted in \autoref{fig:create_datasets}. Whether a secondary log transformation is applied after root transformation or not, if ion curve over time is smoothed or not and the upper clip value. Smoothing is a simple moving average of length 3, repeated twice. Finally, datasets are scaled in 0-255 range and saved as uint8 data type for compression and later use. Integers in 0-255 range is the data range that EfficientNet pretrained models use. This compression effect, from converting float to integers, as well as the data loss from rounding float raw \emph{m/z} values in a previous step, although was not measured (on other than 2D CNN models), wasn't expected to affect performance significant cause data was to noisy and it could have acted as a denoiser. It is though something may worth to be tested in a future work.

\begin{figure}[h]
    
    \centering
    \subfloat[\centering Dataset 1 - Ions 50-54
    plot]{{\includegraphics[width=0.45\textwidth,trim=0 0 0 40, clip]{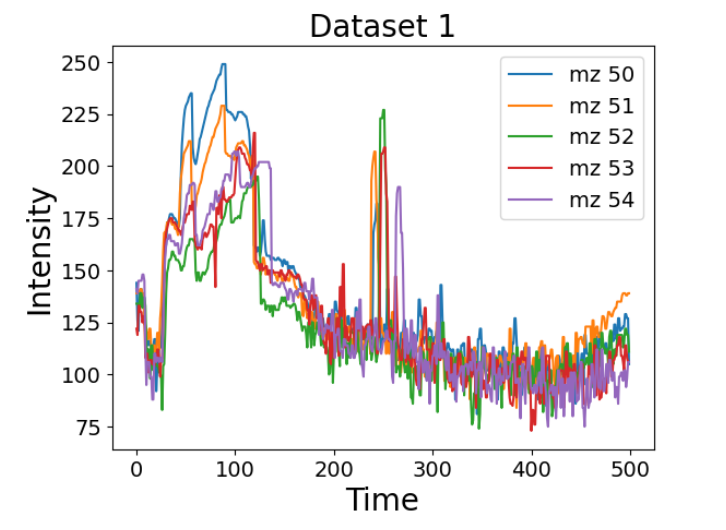} }}%
    \subfloat[\centering Dataset 2 - Ions 50-54 
    plot]{{\includegraphics[width=0.45\textwidth,trim=0 0 0 38, clip]{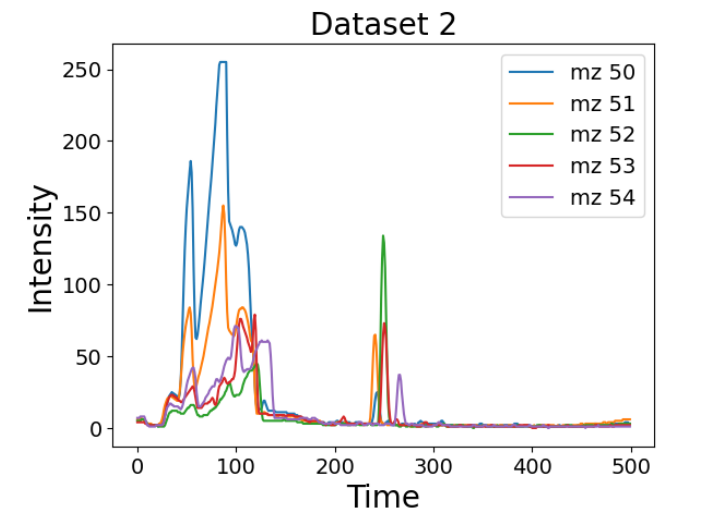} }}%
    \caption{plots of 5 ions for the same sample in two different GC-MS processed datasets}%
    \label{fig:example}%
\end{figure}

Datasets 1 and 1b has gone through both square root and log transformation whether for dataset 2 only square root transformation is used. The only difference between datasets 1 and 1b is that in the first one no smooth function is applied. Plots in \autoref{fig:example} are from the same sample, showing the difference between Dataset 1 and Dataset 2. Dataset 1 has more noise than Dataset 2 but can capture things that escape Dataset 2 focus. As variation is always an important element in machine learning, training models with diverse datasets can increase final performance of ensemble.

After transforming spectrometry values to spectrogram images, our data can be visualized and used by 2D CNN models. In \autoref{fig:SameSample_DifferentDataset}, spectrograms of Dataset 1 and Dataset 1b seem to be much alike but Dataset 2, which is used by neural network models, seems quite different. Horizontal axis represents the time and vertical one the ion (\emph{m/z}). Pixel value is the intensity of an ion at a given timestep. Note that at the bottom there is always a dark stripe of zero values when there are no ions of such \emph{m/z} present in the sample (first row is always zero too, as there is non ion of 0 mass).

All other models, except 2D CNN, are trained with statistical features derived from initial datasets. 
The following statistical feature datasets are used for model training:
\begin{description}
\item[$\bullet$ Statistical features A] dataset comes from Dataset 2, by taking the mean, max and std (standard deviation) features timewise and mean and std ionwise for every sample, all scaled to [0,1] interval. 

\item[$\bullet$ Statistical features B] dataset comes from Dataset 1, by taking the mean, std and skew features timewise and mean ionwise for every sample, scaled to [-0.5,0.5]. 

\item[$\bullet$ Statistical features C] dataset comes from Dataset 1, by taking the mean, max and std features timewise and mean ionwise for every sample, scaled to [0,1]. 

\item[$\bullet$ Statistical features D] dataset comes from Dataset 1b, by taking the mean, max and std features timewise and mean ionwise for every sample, scaled to [0,1]. 
\end{description}

These statistical datasets are much alike. Small differences between them such as  scaling range, statistics used and initial dataset aim to increase individual model performance as different models performed best with a specific type of statistical dataset (as from experimentation). Furthermore, individual models estimations varied more between models, improving final ensemble (average) estimations.

Both datasets were original and hadn't been used before. More info about raw mass spectrometry data can be found on competitions webpages, \href{https://www.drivendata.org/competitions/93/nasa-mars-spectrometry/page/438/}{EGA-MS} and \href{https://www.drivendata.org/competitions/97/nasa-mars-gcms/page/518/}{GC-MS}. Finally, it worth to be mentioned that for both the EGA and GC datasets there was no use of a chromatographic peak detection algorithm, neither any feature creation from peaks. 

\subsection{Metric}
\label{sec:Metric}
Our research and competition's metric is the aggregated logarithmic loss of equation (\ref{eqn:aggLL})
\begin{equation}
\displaystyle
AggLogLoss = -\frac{1}{M*N} \sum\limits_{m=1}^{M}\sum\limits_{n=1}^{N} \left((y_{nm} * \log{ \hat{y}_{nm}})+ ( 1- y_{nm}) * (1-\log{ \hat{y}_{nm}}) \right)  
\label{eqn:aggLL}
\end{equation}

The binary log loss is computed for each possible class and for each sample, and then the mean of results are returned. This is a multilabel classification task with N representing the number of samples and M the number of possible classes. \(y_{nm}\) are the ground truth values and \(\hat{y}_{nm}\) the predicted probabilities. Logarithmic loss provides a steep penalty for predictions that are both confident and wrong.

\subsection{Modelling}
\label{sec:Modelling}
Machine learning models consist of Recurrent Neural Networks (RNNs) using tensorflow and pytorch frameworks for the EGA-MS data and four models (section \ref{sec:GCMS_modelling}), without RNN, but including pretrained CNNs and simple Neural Networks (tensorflow only) for the GC-MS dataset.

\subsubsection{EGA-MS}
\label{sec:EGAMS_modelling}

Modelling consists of 3 types of models, 4 datasets (different bins), 2 levels and 3 repeats. Models used are:

\begin{description}
\item[$\bullet$]  Keras RNN model trained with all 10 classes at final output layer
\item[$\bullet$]  Keras RNN model trained with 1 class at final output layer. (10 models)
\item[$\bullet$]  Pytorch RNN model trained with all 10 classes at final output layer 
\end{description}

\textbf{Model architecture:}. RNN model is of total depth 4 with 1 Bidirectional LSTM and 3 dense layers including the output layer. LSTM layers, \citep{yu2019review}, are best for capturing the time aspect of increasing temperature within recurrent neural networks. In addition, in order to keep a temperature independence aspect, as temperature profile on SAM samples differs from commercial ones, aggregation layers (maxpooling and averagepooling) for every ion are used, representing the maximum and the average value of the ion within it's temperature profile.

Pytorch models weight on final ensemble is relatively small but performed best on out of fold (OOF) SAM samples. Also, within pytorch training pipeline, a weak mixup augmentation is used \citep{zhang2017mixup}, as this increased the OOF score (keras models didn't improve with mixup).

\textbf{Data}:
Keras models were trained with primary and secondary data inputs (2 inputs) whereas pytorch model only with the primary dataset. Four different primary datasets are used, so 4 different trainings are performed. Predictions are averaged over those 4 different runs.

\begin{figure}[h]
    \centering
    \includegraphics[width=0.75\textwidth]{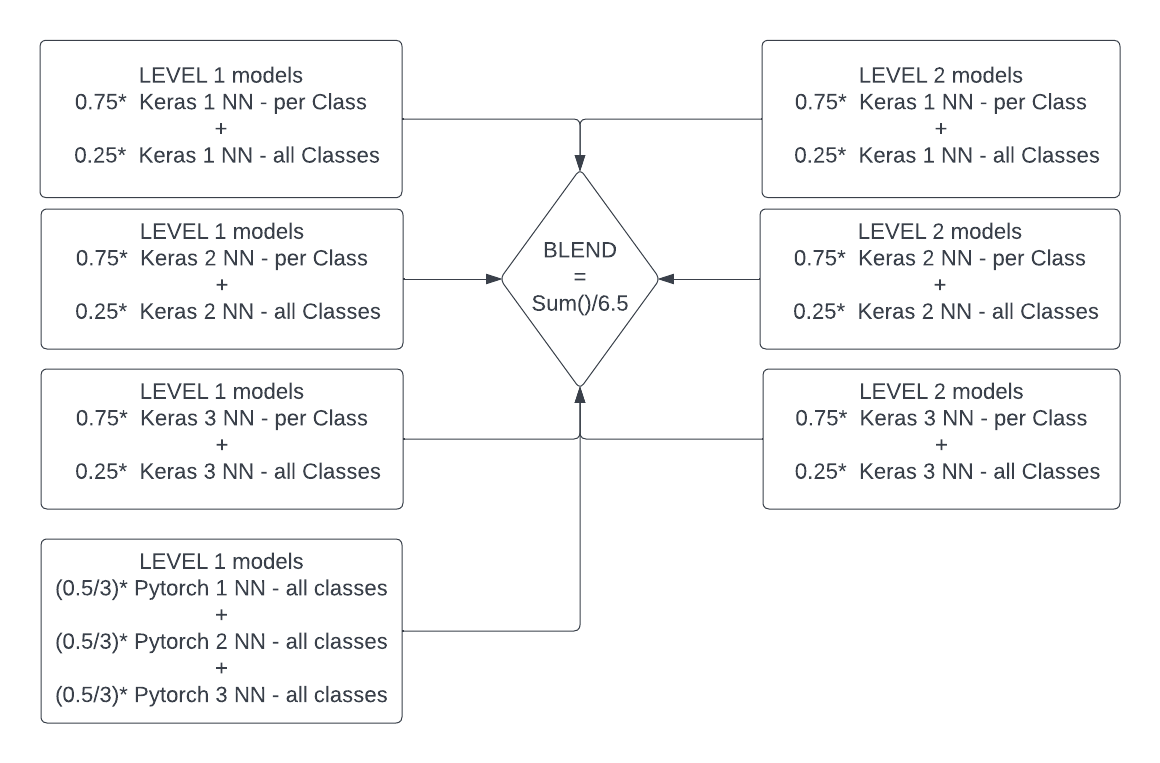}
    \caption{Model ensembling diagram for EGA-MS.}
    \label{fig:Mars_models_blending}
\end{figure}

\textbf{Level 1} models. Every model is trained with every dataset using a 10 fold cross validation stratified split, \citep{stone1974cross}. OOF predictions as well as supplemental data predictions are saved. OOF predictions are the predictions for the training data made by splitting the whole dataset into 10 parts, train using 9 of them and predict on the last one. This is repeated 10 times leaving a different part outside training every time. Out of fold predictions are predictions for the training data from models trained in different part of the data.

\textbf{Level 2} models, Pseudo-labelling. Using average predictions on supplemental data as targets, the training dataset is extended and the whole training is repeated (keras models only). With pseudo-labelling, a widely used machine learning technique, model performance most of the times increases,  \citep{lee2013pseudo}. Here, training additional models including pseudo-labelling on supplemental data where SAM samples are more, improved the OOF results especially for the SAM samples.

\textbf{Repeats}: To improve generalization, every model is run 3 times, each one with a different seed for a different random initial state, using the default keras layers initializers which for the most common dense layer is the glorot uniform method,  \citep{glorot2010understanding}.

\textbf{Ensemble} Final predictions are a weighted average of all models. (\autoref{fig:Mars_models_blending}). With ensembling \citep{dietterich2000ensemble}, a common machine learning technique, averaging models prediction probabilities, increases overall performance.

\subsubsection{GC-MS}
\label{sec:GCMS_modelling}

\begin{figure}[h]
    \centering
    \includegraphics[width=0.75\textwidth]{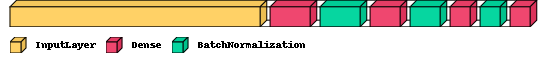}
    \caption{Simple Keras NN architecture.}
    \label{fig:simple_keras_architecture}
\end{figure}

For the GC-MS data, the following models were used:
\begin{description}
\item[$\bullet$]  Simple custom keras over tensorflow artificial neural network (\autoref{fig:simple_keras_architecture})
\item[$\bullet$]  Pretrained on imagenet CNN (EfficientNetB0/1/2 keras over tensorflow), \citep{tan2019efficientnet} 
\item[$\bullet$]  Logistic Regression, \citep{berkson1944application}
\item[$\bullet$]  Ridge classification, \citep{hoerl1970ridge}
\end{description}

\begin{figure}[h]
    \centering
    \includegraphics[width=0.95\textwidth]{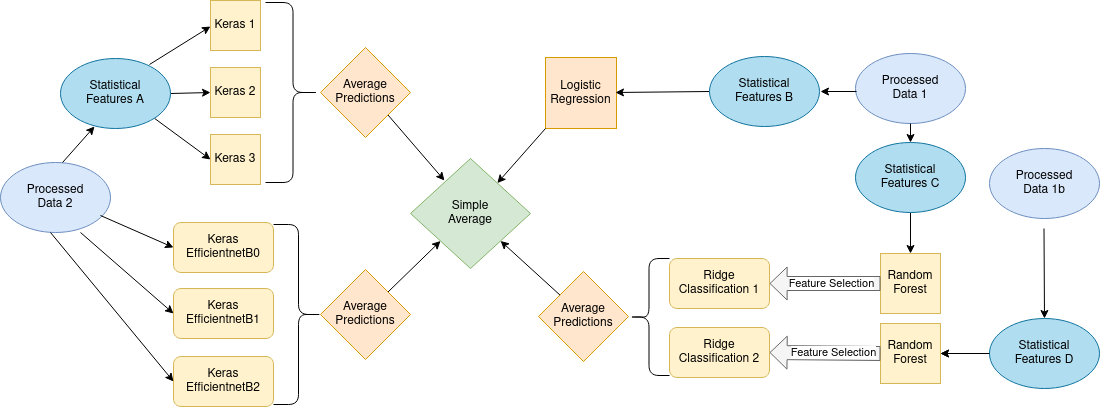}
    \caption{Flowchart of training and inference pipeline for the GC-MS dataset.}
    \label{fig:flowchart_Mars2}
\end{figure}

Random forest models \citep{breiman2001random}, are used for feature selection (dimensionality reduction) purpose only and therefore are not direct part in final ensemble. Selected features used only by ridge classification models  (\autoref{fig:flowchart_Mars2}).

All models except logistic regression was trained multiple times. This helped both individual model score stability and performance. Adding more logistic regression models didn't improve overall performance. Two Ridge classification models were trained, each one with a different training dataset. Both models used random forest to reduce number of features to increase performance and reduce overfitting, as when too many features are used the overfitting probability increases, especially when the number of samples is small \citep{hua2005optimal}. For simple keras models, the same model is used and averaged 3 times and for keras 2D CNN models, 3 different backbones, all belonging in efficientnet family, are used.

Model training was done using a 5 fold cross validation stratified split. A 5 fold instead of the 10 fold split used in EGA-MS was selected for faster experimentation as 2D CNN model training of GC-MS data expected to be computationally heavier than simpler models. Logistic regression and ridge classification models are trained multiple times, one for every of 9 classes (1 output at the time) whereas neural network models trained with all 9 classes together. For neural networks training, the cosine annealing learning rate schedule is used and the best weight of every fold saved for inference. Finally, for 2D CNN model training, light in-layer augmentation is used consisting of one layer for time shift and one layer for random contrast.

\section{Results and discussion}
\label{sec:results and discussion}

Methodologies presented here started with NASA's evolved gas analysis or gas chromatography mass spectrometry labeled data. These data was processed for use by machine learning models and then models were trained using the processed datasets. Finally, trained models was used for inference on the test dataset. Same data processing and machine learning modelling can be applied to any analogue mass spectrometry dataset. Especially for small training datasets, used models and techniques here, can provide valuable insights.

Results and performance evaluation is mandatory for a complete understanding of the subject. \autoref{table:competitions_results} shows results of described methodologies on the unseen test datasets. In bold, the author's performance for both competitions. For the EGA-MS competition, a special prize and key element, was the performance on the SAM testbed samples. The winner of EGA-MS competition performed much better by using pretrained 2D CNN, a technique that lacked from all other competitors on this competition but was present in the following GC-MS competition, empowering top solutions. 

Pretrained CNN worked better for the EGA-MS data than GC-MS. The noisier nature along with the absence of temperature measurements of the GC-MS data constrained the dominance of the 2D CNNs among top solutions and gave room to a variety of other approaches to stand out.

\begin{table}[ht!]
\caption{Competitions results, author's performance in bold}
\label{table:competitions_results}
\centering
\begin{tabular}{l  c  c  c } 
 \hline
 Rank & EGA error (Prec.) & EGA SAM error (Prec.)  & GC error (Prec.) \\ [1ex] 
 \hline
 1 & 0.0920 (0.9483) & 0.1253 (0.5186) & 0.1443 (0.8069) \\ 
 2 & 0.1160 (0.9213) & 0.1257 (0.4238) & 0.1485 (0.8091) \\
 3 & 0.1189 (0.9175) & 0.1320 (0.3558) & \textbf{0.1497 (0.7932)} \\
 4 & 0.1202 \hspace{8ex} & \textbf{0.1346 (0.3378)} & 0.1504 \hspace{8ex} \\ 
 5 & 0.1208 \hspace{8ex} & 0.1359 (0.3652) & 0.1508 \hspace{8ex} \\ 
 6 & 0.1213 \hspace{8ex} &  & 0.1551 \hspace{8ex} \\ 
 7 & \textbf{0.1227} \hspace{8ex} & & 0.1565 \hspace{8ex} \\ 
 \hline
 Benchmark &  0.3242 (0.6378) & 0.3750 (0.1261) & 0.2200 (0.6229) \\ 
 \hline
\end{tabular}

\end{table}

All top scores improved the benchmark a lot for both competitions. For the EGA-MS, the large gap between all other solutions and the first place, fade away on the SAM testbed in regard with competition metric. The error over the SAM samples was just about 10\% higher than commercial samples but Precision (micro average precision) dropped dramatically, over 50\%. Top solution managed to outperform by far all other solution regarding precision on SAM samples too, proving once more the power of pretrained 2D-CNNs on this dataset.
For the GS-MS, though the improvement from benchmark was not that impressive, final scores of top solutions are quite high. More information about winner solution is available at \href{https://drivendata.co/blog/mars-challenge-winners}{(EGA-MS)} and \href{https://drivendata.co/blog/mars-2-gcms-challenge-winners}{(GC-MS)}. 

Competition's metric, average logarithmic loss, is the metric (loss function too) that the models are trained to optimize. If precision was the competition' s metric, precision scores would have been better and logarithmic loss would have been worse. Had precision metric been selected as metric could have led to smaller difference between commercial and SAM instrument precision as SAM's samples weight would have been larger. Selecting logarithmic loss as metric highlighted the importance of the prediction probability been close to the corresponding actual/true value.

Using a square root transformation increased model performance. As can be seen in plots of \autoref{fig:ega_root_noroot}, square root transformation helps weaker lines – ions to emerge and thus to play a more important role within our modelling. At the same time, dominant ions continue to dominate but with less relative power to other ions present. 

For the EGA-MS data using Level 2 models, training models including supplemental data (with level 1 models predictions as labels), may have helped performance on SAM samples but because of the small size of the training dataset, may have hurt the overall performance. As for the following GC-MS competition, pseudo-labelling proved harmful for the OOF score. As of these, modelling pipelines should be simple, without pseudo-labelling, when main annotated data are small.

Method was improved on the GC-MS competition. Including 2D-CNN models as well as extra attention to avoid overfitting on small datasets was essential. Furthermore, the faster data processing due to experience gained in EGA-MS competition left more time available for experimentation. Results presented here are more focused on the GS-MS dataset as corresponding methodology is more accurate and robust.

\begin{table}[ht!]
\caption{Performance of different models on GC-MS data, 5-fold CV scores}
\label{table:model_performance_cv_gcms}
\centering
\begin{tabular}{l  c  c } 
 \hline
 Model Name & Num models & OOF score  \\ [1ex] 
 \hline
 Logistic Regression & 1 & 0.124605  \\ 
 Ridge Classification & 2 & 0.148208 \\
 Simple keras & 3 & 0.116475 \\
 keras EfficientNet(B0/B1/B2) & 3 & 0.122754 \\ 
 \hline
 Ensemble All 4 &   & 0.101963 \\
\hline
\end{tabular}
\end{table}

\autoref{table:model_performance_cv_gcms} shows the OOF performance of every model alone and their ensembling on GC-MS data. The simplest logistic regression model performed very well and the ridge classification models although didn't perform that well alone helped the combined solution. The simple keras neural networks on statistical features performed best but probably were overconfident to the training dataset and finally the pretrained 2D CNNs that won the EGA-MS competition although didn't outperform the other models empowered the ensembling and added reliability and robustness to the final solution.

Although logarithmic loss metric used is one of the most important classification metrics based on probabilities, it is hard to interpret it's values. \autoref{table:ens_performance_perclass_cv_gcms} contains more analytical results using the ensemble OOF predictions of the GC-MS data (labels of the test dataset are not released). For every one of the dataset's compounds there are True Positives (TP), False Positives (FP), True Negatives (TN), False Negatives (FN), Precision (TP / (TP + FP)), Recall (TP / (TP + FN)) and F1 score (2*(Precision*Recall)/(Precision+Recall) reported. As this was a multilabel and not a classification task, making a confusion matrix to understand which class was misclassified with which other class, is impossible. Individual class Precision is quite good as False Positives are not many, but Recall is not that good, especially for the compounds with low occurrences (95\% correlation between Recall and compound counts). High Recall values is due to relatively many False Negatives for those classes. This dataset is not only small but quite unbalanced to. Classes counts range from 26 to 307 and furthermore there are 450 samples that do not contain any class. For every chemical compound, Precision score is higher than Recall, but for samples that contained none of the nine compounds there is a inversion as Recall score is higher than Precision.

The high accuracy of the method on hydrocarbons with a precision score of 0.965 and Recall of 0.909 is impressive, as these are prime targets for the search of organic molecules. To further improve performance for this class a new set of models could be trained with 2 possible classes instead of 9, whether sample is hydrocarbon or not. Furthermore, in cases where reduced False Positives for the hydrocarbon class are necessary, a new metric could be used, such as the Precision score which penalize FPs.

\begin{table}[ht]
\caption{Ensemble's performance (OOF) for every class of GC-MS data}
\label{table:ens_performance_perclass_cv_gcms}
\centering
\begin{tabular} {l l  l  l l l l l}
\hline 
Compound & TP & FP & TN & FN & Prec. & Recall & F1 \\
\hline 
Alcohol & 13 & 0 & 1087 & 21 & 1.0 & 0.382 & 0.553 \\
Aromatic & 55 & 13 & 1005 & 48 & 0.809 & 0.534 & 0.643 \\
Carboxylic\_Acid & 112 & 28 & 928 & 53 & 0.8 & 0.679 & 0.734 \\
Chlorine\_bearing\_compound & 13 & 1 & 1094 & 13 & 0.929 & 0.5 & 0.65 \\
Hydrocarbon & 279 & 10 & 804 & 28 & 0.965 & 0.909 & 0.936 \\
Mineral & 137 & 22 & 919 & 43 & 0.862 & 0.761 & 0.808 \\
Nitrogen\_bearing\_compound & 96 & 29 & 953 & 43 & 0.768 & 0.691 & 0.727 \\
Other\_Oxygen\_bearing\_comp. & 10 & 3 & 1092 & 16 & 0.769 & 0.385 & 0.513 \\
Sulfur\_bearing\_compound  & 18 & 0 & 1085 & 18 & 1.0 & 0.5 & 0.667 \\ \hdashline
None\_of\_9\_compounds(Nans)  & 408 & 130 & 568 & 42 & 0.798 & 0.907 & 0.849 \\ 

\hline 
Average Macro  & & & &  & 0.878 & 0.593 & 0.693 \\
Average Micro  & 733 & 106 & 8967 & 283 & 0.874 & 0.722 &	0.79 \\
\hline 

Average Macro with Nans  & & & &  & 0.87 & 0.625 & 0.708 \\
Average Micro with Nans  & 1141 & 209 & 9535 & 325 & 0.845 &  0.778 & 0.81 \\
\hline 

\end{tabular}

\end{table}

Presented results demonstrated the capabilities of artificial intelligence models on mass spectrometry data for detecting the presence of specific chemical compounds in geological material samples of planet Mars analogues. Whether these results indicate past habitability or not is for the experts to decide. Current research could provide valuable insights for future designs of planetary mission instruments performing in situ analysis on Mars or any other terrestrial body. Furthermore, could be used to help the work of flight data analysis teams by pre-assessing the data content before in-depth treatment.

\subsection{Generalization and Efficiency of used machine learning models}
\label{sec:Generalization and Efficiency}

Models trained on small datasets as this one, can overfit or be overconfident and perform well only with very similar distributions datasets. Towards a robust generalized approach the following used here models are suggested: 

\begin{itemize}
\item[--] Ridge classification models which use l2 regularization that can lower overfitting. Performance and generalizations was enhanced for these models when random forest applied first for most important feature selection.
\item[--] Logistic regression which only needs 1 parameter to be tuned (C), which was kept constant over all folds and all classes
\item[--] Pretrained on imagenet models which have as a training starting point the finishing point of training on the imagenet dataset (had already seen many, completely different images). Selected pretrained CNNs EfficientNetB0, EfficientNetB1 and EfficientNetB2 are both fast and accurate. 
\end{itemize}

GC-MS modelling also included Simple NN which though performed much better on OOF, may have overfitted a little more than the other models and therefor are underweighted in the final ensemble. All models are equal weighted despite the fact that CV indicated an increased weight for simple NN, 2.5 times more than the rest models (\autoref{fig:flowchart_Mars2}).

Apart from robustness, inference speed without loosing accuracy was also a concern, as these models could be used in space missions. At inference time, 3 out of 4 models  were very fast using an "Intel(R) Xeon(R) CPU @ 2.20GHz" with 4 cores CPU and even 2D CNN models can be used with CPU quite fast. GPU or any other accelerator is not necessary for inference. Creating all necessary datasets takes 64 minutes for all 1584 train and test samples and that's an average of 2.42 seconds per sample (single thread). Also for the 3 first models an average inference time of 0.025 seconds per sample is needed and heavier 2D CNN models take 0.3 seconds per sample. Inference time can be reduced by 5 times if 5-fold training is replaced by a single full training and moreover the dataset can be created using multithreading (for supported CPUs). Pipeline as it is needs less than 3 seconds for a single sample to go from raw spectrometry data to classification outcome using python, a scripting language good for prototyping but relative slow for production. Applications written in C or C++, as those often used in flight missions, will be a few times faster.

Adding an automation system to use artificial intelligence models on mass spectrometry data to a planetary rover is achievable for forthcoming missions. Data processing and models presented can provide valuable insights in this endeavor. Processing power in future rovers can be improved following technological advancements. Current approach is not high resources demanding and can be used with slower CPUs.

\subsection{Things that didn't work and perspectives}
\label{sec:Also tried}
Models and techniques that were tested but performed worse and didn't make it to the final ensemble can be seen at \autoref{table:tried}.

\begin{table}[ht!]
\caption{Thing that tried but didn't make it to final ensemble.}
\label{table:tried}
\centering
\begin{tabular}{l  l } 
 \hline
 EGA-MS & GC-MS  \\ [1ex] 
 \hline
    lightgbm (model) & lightgbm \\
    tabnet (model) & xgboost (model) \\
    SVM (model)  & SVM (good score alone) \\
   & NN with one label output at the time \\    
   & 1 dimensional CNN (Conv1D) \\
   & recurrent neural networks \\
   & transformers (NN) \\    
   & pseudo-labelling  \\ 
   & sample mixup augmentation \\
   & using less \emph{m/z} ions\\
   & using less or more timesteps\\
\hline
\end{tabular}

\end{table}

Many models and techniques were tried along the way but was not used at the end. Models that on early stages of experimentation seemed promising dropped out gradually as improvements in neural networks performance made them redundant. For the GC-MS data more experiments than EGA-MS was done as experience on preceded EGA-MS data gave a head start and left more time available for more advanced experimentation. Furthermore a 5-fold instead a 10-fold validation increased experiment's speed, though using more folds should be safer to be applied on small datasets. Pseudo-labelling shouldn't had been used cause increased modelling complexity and decreased performance for the GC-MS data while for the EGA-MS the gain was minor. Mixup augmentation added more noise in an already noisy dataset and worsen results. Finally, making spectrograms with less ions or with less or more timesteps, a different shape of spectrogram, also made results worse.

These are two interesting and original datasets. For continue working, it would be of great benefit of the results if more data were available for training, as both datasets are small. Models that didn't work before may work with additional data and/or models that was finally used may not be required any more. Continue working with these data could include:
\begin{itemize}
\item[--] Use (more) pretrain CNN architectures,
\item[--] Try pretrain CNN with pytorch,
\item[--] More experimentation using pytorch framework,
\item[--] Use peak detection to extract features,
\item[--] Remove background noise by subtracting the intensity value that immediately precedes or follows a peak (competition organizer suggestion).
\item[--] Try float raw \emph{m/z} values or other rounding options,
\item[--] Rethinking all things that didn't work.

\end{itemize}

\section{Conclusions}
\label{sec:conclusions}
Machine learning use computational models to "learn" information directly from data without relying on predetermined equations and turns raw data to actionable insights. Using various artificial intelligence models on mass spectrometry data, quite accurate results within a short time are acquired. This can be of great importance in future missions for in-flight processing of mass spectrometry data. To increase accuracy, root transformation of mass spectrometry intensity/abundance values was important. Furthermore, created 2D spectrograms allowed the use of pretrained CNNs which performed exceptionally well. Generalization is a key element when working with small datasets. Using appropriate machine learning models, model ensembling and right training procedures was crucial to minimize overfitting and overconfidence. For more accurate results, for both tasks,  more data are needed and upon higher data availability, results will be improved. Although models trained mostly with data from commercial instruments, predictions performance on SAM EGA-MS was promising. Evolved Gas Analysis as well as Gas Chromatography mass spectrometry combined with machine learning are both valuable in the analysis of sediments from planet Mars and in the same way from any other terrestrial body of our solar system. 
Mass spectrometry data analysis with artificial intelligence can be run on the edge effectively and future missions can take advantage of it.

\section*{Declaration of Competing Interest}
The author declares that has no known competing financial interests or personal relationships that could have appeared to influence the work reported in this paper.

\section*{Acknowledgment}
\label{sec:Acknowledgment}
NASA provided support for the development of SAM. The datasets for these 2 challenges were provided by NASA Goddard Space Flight Center and NASA Johnson Space Center. They have been collected by the Sample Analysis at Mars (SAM) scientists and specifically processed for these challenges with the help of the scientists from NASA: Doug Archer, Charles Malespin, Caroline Freissinet, Stephanie Getty, Luoth Chou, Eric Lyness, and Victoria Da Poian, and the DrivenData team. Data from all SAM experiments are archived in the Planetary Data System (pds.nasa.gov).

\section*{Data availability}
\label{sec:Data availability}
Both datasets were also published after competitions ended at \cite{website:opendata.awsNASA-EGAMS} and at \cite{website:opendata.awsNASA-GCMS}.

\bibliographystyle{cas-model2-names}

\bibliography{refs}

\begin{thebibliography}{29}
\expandafter\ifx\csname natexlab\endcsname\relax\def\natexlab#1{#1}\fi
\providecommand{\url}[1]{\texttt{#1}}
\providecommand{\href}[2]{#2}
\providecommand{\path}[1]{#1}
\providecommand{\DOIprefix}{doi:}
\providecommand{\ArXivprefix}{arXiv:}
\providecommand{\URLprefix}{URL: }
\providecommand{\Pubmedprefix}{pmid:}
\providecommand{\doi}[1]{\href{http://dx.doi.org/#1}{\path{#1}}}
\providecommand{\Pubmed}[1]{\href{pmid:#1}{\path{#1}}}
\providecommand{\bibinfo}[2]{#2}
\ifx\xfnm\relax \def\xfnm[#1]{\unskip,\space#1}\fi
\bibitem[{Aghili et~al.(2022)Aghili, Rasekh, Karami, Azizi and Gancarz}]{aghili2022detection}
\bibinfo{author}{Aghili, N.S.}, \bibinfo{author}{Rasekh, M.}, \bibinfo{author}{Karami, H.}, \bibinfo{author}{Azizi, V.}, \bibinfo{author}{Gancarz, M.}, \bibinfo{year}{2022}.
\newblock \bibinfo{title}{Detection of fraud in sesame oil with the help of artificial intelligence combined with chemometrics methods and chemical compounds characterization by gas chromatography--mass spectrometry}.
\newblock \bibinfo{journal}{Lwt} \bibinfo{volume}{167}, \bibinfo{pages}{113863}.
\bibitem[{Berkson(1944)}]{berkson1944application}
\bibinfo{author}{Berkson, J.}, \bibinfo{year}{1944}.
\newblock \bibinfo{title}{Application of the logistic function to bio-assay}.
\newblock \bibinfo{journal}{Journal of the American statistical association} \bibinfo{volume}{39}, \bibinfo{pages}{357--365}.
\bibitem[{Bertaux et~al.(2006)Bertaux, Korablev, Perrier, Quemerais, Montmessin, Leblanc, Lebonnois, Rannou, Lef{\`e}vre, Forget et~al.}]{bertaux2006spicam}
\bibinfo{author}{Bertaux, J.L.}, \bibinfo{author}{Korablev, O.}, \bibinfo{author}{Perrier, S.}, \bibinfo{author}{Quemerais, E.}, \bibinfo{author}{Montmessin, F.}, \bibinfo{author}{Leblanc, F.}, \bibinfo{author}{Lebonnois, S.}, \bibinfo{author}{Rannou, P.}, \bibinfo{author}{Lef{\`e}vre, F.}, \bibinfo{author}{Forget, F.}, et~al., \bibinfo{year}{2006}.
\newblock \bibinfo{title}{Spicam on mars express: Observing modes and overview of uv spectrometer data and scientific results}.
\newblock \bibinfo{journal}{Journal of Geophysical Research: Planets} \bibinfo{volume}{111}.
\bibitem[{Breiman(2001)}]{breiman2001random}
\bibinfo{author}{Breiman, L.}, \bibinfo{year}{2001}.
\newblock \bibinfo{title}{Random forests}.
\newblock \bibinfo{journal}{Machine learning} \bibinfo{volume}{45}, \bibinfo{pages}{5--32}.
\bibitem[{Chou et~al.(2021)Chou, Mahaffy, Trainer, Eigenbrode, Arevalo, Brinckerhoff, Getty, Grefenstette, Da~Poian, Fricke et~al.}]{chou2021planetary}
\bibinfo{author}{Chou, L.}, \bibinfo{author}{Mahaffy, P.}, \bibinfo{author}{Trainer, M.}, \bibinfo{author}{Eigenbrode, J.}, \bibinfo{author}{Arevalo, R.}, \bibinfo{author}{Brinckerhoff, W.}, \bibinfo{author}{Getty, S.}, \bibinfo{author}{Grefenstette, N.}, \bibinfo{author}{Da~Poian, V.}, \bibinfo{author}{Fricke, G.M.}, et~al., \bibinfo{year}{2021}.
\newblock \bibinfo{title}{Planetary mass spectrometry for agnostic life detection in the solar system}.
\newblock \bibinfo{journal}{Frontiers in Astronomy and Space Sciences} \bibinfo{volume}{8}, \bibinfo{pages}{755100}.
\bibitem[{Clark et~al.(2020)Clark, Archer, Gruener, Ming, Tu, Niles and Mertzman}]{clark2020jsc}
\bibinfo{author}{Clark, J.}, \bibinfo{author}{Archer, P.}, \bibinfo{author}{Gruener, J.}, \bibinfo{author}{Ming, D.}, \bibinfo{author}{Tu, V.}, \bibinfo{author}{Niles, P.}, \bibinfo{author}{Mertzman, S.}, \bibinfo{year}{2020}.
\newblock \bibinfo{title}{Jsc-rocknest: A large-scale mojave mars simulant (mms) based soil simulant for in-situ resource utilization water-extraction studies}.
\newblock \bibinfo{journal}{Icarus} \bibinfo{volume}{351}, \bibinfo{pages}{113936}.
\bibitem[{Clegg et~al.(2014)Clegg, Wiens, Misra, Sharma, Lambert, Bender, Newell, Nowak-Lovato, Smrekar, Dyar et~al.}]{clegg2014planetary}
\bibinfo{author}{Clegg, S.M.}, \bibinfo{author}{Wiens, R.}, \bibinfo{author}{Misra, A.K.}, \bibinfo{author}{Sharma, S.K.}, \bibinfo{author}{Lambert, J.}, \bibinfo{author}{Bender, S.}, \bibinfo{author}{Newell, R.}, \bibinfo{author}{Nowak-Lovato, K.}, \bibinfo{author}{Smrekar, S.}, \bibinfo{author}{Dyar, M.D.}, et~al., \bibinfo{year}{2014}.
\newblock \bibinfo{title}{Planetary geochemical investigations using raman and laser-induced breakdown spectroscopy}.
\newblock \bibinfo{journal}{Applied spectroscopy} \bibinfo{volume}{68}, \bibinfo{pages}{925--936}.
\bibitem[{Dietterich(2000)}]{dietterich2000ensemble}
\bibinfo{author}{Dietterich, T.G.}, \bibinfo{year}{2000}.
\newblock \bibinfo{title}{Ensemble methods in machine learning}, in: \bibinfo{booktitle}{International workshop on multiple classifier systems}, \bibinfo{organization}{Springer}. pp. \bibinfo{pages}{1--15}.
\bibitem[{Fouchet et~al.(2022)Fouchet, Reess, Montmessin, Hassen-Khodja, Nguyen-Tuong, Humeau, Jacquinod, Lapauw, Parisot, Bonafous et~al.}]{fouchet2022supercam}
\bibinfo{author}{Fouchet, T.}, \bibinfo{author}{Reess, J.M.}, \bibinfo{author}{Montmessin, F.}, \bibinfo{author}{Hassen-Khodja, R.}, \bibinfo{author}{Nguyen-Tuong, N.}, \bibinfo{author}{Humeau, O.}, \bibinfo{author}{Jacquinod, S.}, \bibinfo{author}{Lapauw, L.}, \bibinfo{author}{Parisot, J.}, \bibinfo{author}{Bonafous, M.}, et~al., \bibinfo{year}{2022}.
\newblock \bibinfo{title}{The supercam infrared spectrometer for the perseverance rover of the mars2020 mission}.
\newblock \bibinfo{journal}{Icarus} \bibinfo{volume}{373}, \bibinfo{pages}{114773}.
\bibitem[{Glorot and Bengio(2010)}]{glorot2010understanding}
\bibinfo{author}{Glorot, X.}, \bibinfo{author}{Bengio, Y.}, \bibinfo{year}{2010}.
\newblock \bibinfo{title}{Understanding the difficulty of training deep feedforward neural networks}, in: \bibinfo{booktitle}{Proceedings of the thirteenth international conference on artificial intelligence and statistics}, \bibinfo{organization}{JMLR Workshop and Conference Proceedings}. pp. \bibinfo{pages}{249--256}.
\bibitem[{Hoerl and Kennard(1970)}]{hoerl1970ridge}
\bibinfo{author}{Hoerl, A.E.}, \bibinfo{author}{Kennard, R.W.}, \bibinfo{year}{1970}.
\newblock \bibinfo{title}{Ridge regression: Biased estimation for nonorthogonal problems}.
\newblock \bibinfo{journal}{Technometrics} \bibinfo{volume}{12}, \bibinfo{pages}{55--67}.
\bibitem[{Hua et~al.(2005)Hua, Xiong, Lowey, Suh and Dougherty}]{hua2005optimal}
\bibinfo{author}{Hua, J.}, \bibinfo{author}{Xiong, Z.}, \bibinfo{author}{Lowey, J.}, \bibinfo{author}{Suh, E.}, \bibinfo{author}{Dougherty, E.R.}, \bibinfo{year}{2005}.
\newblock \bibinfo{title}{Optimal number of features as a function of sample size for various classification rules}.
\newblock \bibinfo{journal}{Bioinformatics} \bibinfo{volume}{21}, \bibinfo{pages}{1509--1515}.
\bibitem[{Lee et~al.(2013)}]{lee2013pseudo}
\bibinfo{author}{Lee, D.H.}, et~al., \bibinfo{year}{2013}.
\newblock \bibinfo{title}{Pseudo-label: The simple and efficient semi-supervised learning method for deep neural networks}, in: \bibinfo{booktitle}{Workshop on challenges in representation learning, ICML}, p. \bibinfo{pages}{896}.
\bibitem[{Millan et~al.(2016)Millan, Szopa, Buch, Coll, Glavin, Freissinet, Navarro-Gonz{\'a}lez, Fran{\c{c}}ois, Coscia, Bonnet et~al.}]{millan2016situ}
\bibinfo{author}{Millan, M.}, \bibinfo{author}{Szopa, C.}, \bibinfo{author}{Buch, A.}, \bibinfo{author}{Coll, P.}, \bibinfo{author}{Glavin, D.P.}, \bibinfo{author}{Freissinet, C.}, \bibinfo{author}{Navarro-Gonz{\'a}lez, R.}, \bibinfo{author}{Fran{\c{c}}ois, P.}, \bibinfo{author}{Coscia, D.}, \bibinfo{author}{Bonnet, J.Y.}, et~al., \bibinfo{year}{2016}.
\newblock \bibinfo{title}{In situ analysis of martian regolith with the sam experiment during the first mars year of the msl mission: Identification of organic molecules by gas chromatography from laboratory measurements}.
\newblock \bibinfo{journal}{Planetary and Space Science} \bibinfo{volume}{129}, \bibinfo{pages}{88--102}.
\bibitem[{Nardella et~al.(2021)Nardella, Bellavia, Mattonai and Ribechini}]{nardella2021co}
\bibinfo{author}{Nardella, F.}, \bibinfo{author}{Bellavia, S.}, \bibinfo{author}{Mattonai, M.}, \bibinfo{author}{Ribechini, E.}, \bibinfo{year}{2021}.
\newblock \bibinfo{title}{Co-pyrolysis of wood and plastic: Evaluation of synergistic effects and kinetic data by evolved gas analysis-mass spectrometry (ega-ms)}.
\newblock \bibinfo{journal}{Journal of Analytical and Applied Pyrolysis} \bibinfo{volume}{159}, \bibinfo{pages}{105308}.
\bibitem[{NASA(2023a)}]{website:opendata.awsNASA-GCMS}
\bibinfo{author}{NASA}, \bibinfo{year}{2023}a.
\newblock \bibinfo{title}{Mars spectrometry 2: Gas chromatography for the sample analysis at mars data (sam) instrument}.
\newblock \URLprefix \url{https://registry.opendata.aws/nasa-gcms/}.
\bibitem[{NASA(2023b)}]{website:opendata.awsNASA-EGAMS}
\bibinfo{author}{NASA}, \bibinfo{year}{2023}b.
\newblock \bibinfo{title}{Mars spectrometry: Detect evidence for past habitability}.
\newblock \URLprefix \url{https://registry.opendata.aws/nasa-ega/}.
\bibitem[{Ono et~al.(2022)Ono, Rothrock, Iwashita, Higa, Timmaraju, Sahnoune, Qiu, Islam, Didier, Laporte et~al.}]{ono2022machine}
\bibinfo{author}{Ono, M.}, \bibinfo{author}{Rothrock, B.}, \bibinfo{author}{Iwashita, Y.}, \bibinfo{author}{Higa, S.}, \bibinfo{author}{Timmaraju, V.}, \bibinfo{author}{Sahnoune, S.}, \bibinfo{author}{Qiu, D.}, \bibinfo{author}{Islam, T.}, \bibinfo{author}{Didier, A.}, \bibinfo{author}{Laporte, C.}, et~al., \bibinfo{year}{2022}.
\newblock \bibinfo{title}{Machine learning for planetary rovers}, in: \bibinfo{booktitle}{Machine Learning for Planetary Science}. \bibinfo{publisher}{Elsevier}, pp. \bibinfo{pages}{169--191}.
\bibitem[{Pastor et~al.(2022)Pastor, Ili{\'c}, Koji{\'c}, A{\v{c}}anski and Vuji{\'c}}]{pastor2022classification}
\bibinfo{author}{Pastor, K.}, \bibinfo{author}{Ili{\'c}, M.}, \bibinfo{author}{Koji{\'c}, J.}, \bibinfo{author}{A{\v{c}}anski, M.}, \bibinfo{author}{Vuji{\'c}, D.}, \bibinfo{year}{2022}.
\newblock \bibinfo{title}{Classification of cereal flour by gas chromatography--mass spectrometry (gc-ms) liposoluble fingerprints and automated machine learning}.
\newblock \bibinfo{journal}{Analytical Letters} \bibinfo{volume}{55}, \bibinfo{pages}{2220--2226}.
\bibitem[{Slingerland et~al.(2022)Slingerland, Perry, Kaufman, Bycroft, Linstead, Mandrake, Doran, Goel, Feather, Fesq et~al.}]{slingerland2022adapting}
\bibinfo{author}{Slingerland, P.}, \bibinfo{author}{Perry, L.}, \bibinfo{author}{Kaufman, J.}, \bibinfo{author}{Bycroft, B.}, \bibinfo{author}{Linstead, E.}, \bibinfo{author}{Mandrake, L.}, \bibinfo{author}{Doran, G.}, \bibinfo{author}{Goel, A.}, \bibinfo{author}{Feather, M.S.}, \bibinfo{author}{Fesq, L.}, et~al., \bibinfo{year}{2022}.
\newblock \bibinfo{title}{Adapting a trusted ai framework to space mission autonomy}, in: \bibinfo{booktitle}{2022 IEEE Aerospace Conference (AERO)}, \bibinfo{organization}{IEEE}. pp. \bibinfo{pages}{1--20}.
\bibitem[{Stone(1974)}]{stone1974cross}
\bibinfo{author}{Stone, M.}, \bibinfo{year}{1974}.
\newblock \bibinfo{title}{Cross-validatory choice and assessment of statistical predictions}.
\newblock \bibinfo{journal}{Journal of the royal statistical society: Series B (Methodological)} \bibinfo{volume}{36}, \bibinfo{pages}{111--133}.
\bibitem[{Tan and Kerr(2018)}]{tan2018determining}
\bibinfo{author}{Tan, J.}, \bibinfo{author}{Kerr, W.L.}, \bibinfo{year}{2018}.
\newblock \bibinfo{title}{Determining degree of roasting in cocoa beans by artificial neural network (ann)-based electronic nose system and gas chromatography/mass spectrometry (gc/ms)}.
\newblock \bibinfo{journal}{Journal of the Science of Food and Agriculture} \bibinfo{volume}{98}, \bibinfo{pages}{3851--3859}.
\bibitem[{Tan and Le(2019)}]{tan2019efficientnet}
\bibinfo{author}{Tan, M.}, \bibinfo{author}{Le, Q.}, \bibinfo{year}{2019}.
\newblock \bibinfo{title}{Efficientnet: Rethinking model scaling for convolutional neural networks}, in: \bibinfo{booktitle}{International conference on machine learning}, \bibinfo{organization}{PMLR}. pp. \bibinfo{pages}{6105--6114}.
\bibitem[{Varatharajan et~al.(2021)Varatharajan, Angerhausen, Antoniadou, Bickel, D'Amore, Faragalli, L{\'o}pez-Francos, Maiti, Potter, Shneider et~al.}]{varatharajan2021artificial}
\bibinfo{author}{Varatharajan, I.}, \bibinfo{author}{Angerhausen, D.}, \bibinfo{author}{Antoniadou, E.}, \bibinfo{author}{Bickel, V.}, \bibinfo{author}{D'Amore, M.}, \bibinfo{author}{Faragalli, M.}, \bibinfo{author}{L{\'o}pez-Francos, I.}, \bibinfo{author}{Maiti, A.}, \bibinfo{author}{Potter, R.W.}, \bibinfo{author}{Shneider, C.}, et~al., \bibinfo{year}{2021}.
\newblock \bibinfo{title}{Artificial intelligence for the advancement of lunar and planetary science and exploration}.
\newblock \bibinfo{journal}{Bulletin of the American Astronomical Society} \bibinfo{volume}{53}, \bibinfo{pages}{222}.
\bibitem[{Verchovsky et~al.(2020)Verchovsky, Anand, Barber, Sheridan and Morgan}]{verchovsky2020quantitative}
\bibinfo{author}{Verchovsky, A.}, \bibinfo{author}{Anand, M.}, \bibinfo{author}{Barber, S.}, \bibinfo{author}{Sheridan, S.}, \bibinfo{author}{Morgan, G.}, \bibinfo{year}{2020}.
\newblock \bibinfo{title}{A quantitative evolved gas analysis for extra-terrestrial samples}.
\newblock \bibinfo{journal}{Planetary and Space Science} \bibinfo{volume}{181}, \bibinfo{pages}{104830}.
\bibitem[{Wong et~al.(2022)Wong, Franz, Clark, McAdam, Lewis, Millan, Ming, Gomez, Clark, Eigenbrode et~al.}]{wong2022oxidized}
\bibinfo{author}{Wong, G.M.}, \bibinfo{author}{Franz, H.B.}, \bibinfo{author}{Clark, J.V.}, \bibinfo{author}{McAdam, A.C.}, \bibinfo{author}{Lewis, J.M.}, \bibinfo{author}{Millan, M.}, \bibinfo{author}{Ming, D.W.}, \bibinfo{author}{Gomez, F.}, \bibinfo{author}{Clark, B.}, \bibinfo{author}{Eigenbrode, J.L.}, et~al., \bibinfo{year}{2022}.
\newblock \bibinfo{title}{Oxidized and reduced sulfur observed by the sample analysis at mars (sam) instrument suite on the curiosity rover within the glen torridon region at gale crater, mars}.
\newblock \bibinfo{journal}{Journal of Geophysical Research: Planets} \bibinfo{volume}{127}, \bibinfo{pages}{e2021JE007084}.
\bibitem[{Yu et~al.(2019)Yu, Si, Hu and Zhang}]{yu2019review}
\bibinfo{author}{Yu, Y.}, \bibinfo{author}{Si, X.}, \bibinfo{author}{Hu, C.}, \bibinfo{author}{Zhang, J.}, \bibinfo{year}{2019}.
\newblock \bibinfo{title}{A review of recurrent neural networks: Lstm cells and network architectures}.
\newblock \bibinfo{journal}{Neural computation} \bibinfo{volume}{31}, \bibinfo{pages}{1235--1270}.
\bibitem[{Zhang et~al.(2017)Zhang, Cisse, Dauphin and Lopez-Paz}]{zhang2017mixup}
\bibinfo{author}{Zhang, H.}, \bibinfo{author}{Cisse, M.}, \bibinfo{author}{Dauphin, Y.N.}, \bibinfo{author}{Lopez-Paz, D.}, \bibinfo{year}{2017}.
\newblock \bibinfo{title}{mixup: Beyond empirical risk minimization}.
\newblock \bibinfo{journal}{arXiv preprint arXiv:1710.09412} .
\bibitem[{Zumaquero~Silvero et~al.(2020)Zumaquero~Silvero, Gilabert~Albiol, D{\'\i}az-Canales, Ventura~Vaquer and G{\'o}mez-Tena}]{zumaquero2020application}
\bibinfo{author}{Zumaquero~Silvero, E.}, \bibinfo{author}{Gilabert~Albiol, J.}, \bibinfo{author}{D{\'\i}az-Canales, E.M.}, \bibinfo{author}{Ventura~Vaquer, M.J.}, \bibinfo{author}{G{\'o}mez-Tena, M.P.}, \bibinfo{year}{2020}.
\newblock \bibinfo{title}{Application of evolved gas analysis technique for speciation of minor minerals in clays}.
\newblock \bibinfo{journal}{Minerals} \bibinfo{volume}{10}, \bibinfo{pages}{824}.

\end{thebibliography}

\end{document}